# Foveated Time Stretch

Jacky C.K. Chan, and Bahram Jalali

Department of Electrical and Computer Engineering, UCLA, Los Angeles, CA, USA, 90095

**Abstract**

*Given prior knowledge of the spectral statistics, the SNR of optical waveforms can be manipulated in a reversible manner. We introduce this concept and discuss its potential application to encryption and context-aware detection of weak signals in noisy environments. The same technology performs single shot sparse sampling for optical data compression.*

## I. INTRODUCTION

Photonic hardware accelerators (PHAs) comprise a new category of data processing engines, in which analog optical pre-processing is performed preceding optical-to-electrical conversion and digital processing to alleviate the burden on the electronics. This vision enables the acquisition and digital processing of waveforms that are much faster than what is possible with conventional electronics [1], [2]. An early example of PHAs, the photonic time stretch data acquisition [3], [4] has led to the discovery of optical rogue waves [5], and the observation of relativistic electron microstructures in synchrotron radiation [6]. It has made it possible to see the birth of laser modelocking [7] and internal motion of soliton molecules [8], ultrafast imaging [9] and Raman spectroscopy [10]. In combination with artificial intelligence, time stretch imaging has led to label-free detection of cancer cells in blood with record specificity [11].

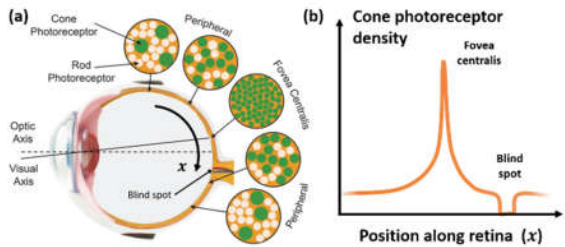

Figure 1: (a) Cross-section of the human eye. The cone photoreceptor density varies along the surface of the retina. (b) The fovea centralis is a small area near the centre of the retina, corresponding to cells that capture light for the central field of vision, in which nearly 50% of densely packed cone photoreceptor cells are located. The eye achieves image acquisition with non-uniform resolution in the visual field by spatially varying the cone cell density.

The warped time stretch is a specific class of PHAs that leads to nonuniform sampling of the waveform in the same spirit as nonuniform sampling performed by the fovea centralis in the eye [12]. Similar to linear time stretch, foveated stretch is realized using dispersive elements, but here the dispersion is tailored to map the spectrum into time in a signal-dependent manner [13]. This represents a new type of sparse sampling that operates in real-time (in contrast to conventional sparse sampling that requires multiple measurements followed by iterative algorithms for signal reconstruction). The design of the dispersion profile is tailored to the time-averaged spectral statistics. It is open loop and does not require instantaneous adaptivity. Real-time optical data compression has been demonstrated with this technique to solve the big data predicament caused by the high throughput of time stretch instruments [14]. This has motivated similar follow-on works by a growing number of researchers [15]–[20].

The term "foveated sampling" comes from the "fovea centralis", a region near the centre of the retina at which the highest cell density occurs. The spatially-varying photoreceptor cell density provides non-uniform spatial sampling and enables enhanced image resolution in central vision with lower resolution in the peripheral vision where high resolution is not necessary.

## II. APPROACH

Our goal is to implement the same function for sampling of the spectrum of ultrafast waveforms, and to do so in real-time. Normally this would require a sampler with a sample rate that adapts to the instantaneous behaviour of the input signal in real-time. However, this is not possible because the time scales of optical signals are in the picosecond range or less, which is much shorter than the response time found in electronic circuits. In addition, the sampling rate would have to be synchronized with the variations in input signal.

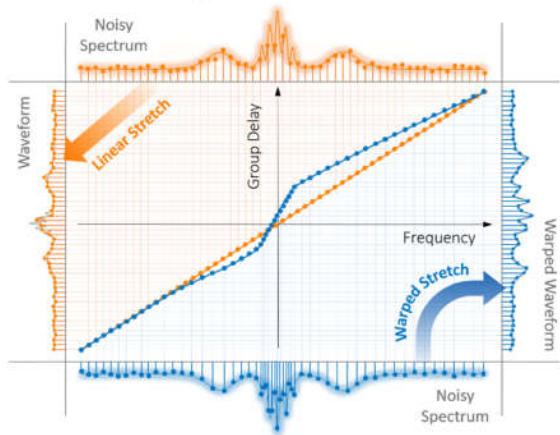

Fig. 1. Warped time stretching allows for context-aware resampling of the spectrum both to slow down fast portions of the signal, while also SNR boosting of the weaker regions. Adapted from [13].

A better approach that avoids the speed and synchronization issues is to perform warped time-stretching followed by uniform sampling. The first step is a spectrotemporal operation in which the spectrum of the broadband optical signal is non-uniformly mapped to a time scale that is slow enough to digitized. This is followed by uniform sampling in time domain performed by an analog-to-digital converter. When combined, this effectively achieves foveated spectral sampling at ultra-high speeds and in real-time. As we demonstrate below, this technique leads to efficient allocation of samples based on the information density of the spectrum.

Our previous theoretical work on warped time stretch provided a systematic approach to analyze the effect of a tailored group delay dispersion profile on an optical signal [5]. The time-bandwidth product of the signal can then be minimized when the group delay profile is designed specifically to the local density of spectral features. Moreover, with an appropriate basis decomposition of the dispersion function, complex stretch operations can be synthesized, opening a new pathway to analog optical computing [6].

Here we show that with proper design of the dispersive profile (group delay vs. optical frequency) both the signal and the noise power can be tailored. We show that how this opens a valuable pathway to SNR engineering for wideband optical data. We also discuss application to optical data encryption in which a signal is hidden under noise.

## III. MATHEMATICAL ANALYSIS

In dispersive optical systems, the output photocurrent $i_{out}(t)$ following photodetection is described as:

$$i_{out}(t) = \frac{c\epsilon_0 n r_d}{4\pi} [\tau_g'(\Omega_s)]^{-1} |\tilde{E}_{in}(\Omega_s)|^2 \quad (1)$$

where $r_d$ is the detector responsivity, $c$ is the speed of light, $\epsilon_0$ is the vacuum permittivity, $n$ is the refractive index of the medium, and $E_{out}(t)$ and $\tilde{E}_{in}(\Omega_s)$ are the electric field and its spectrum, respectively. Implicit in the equation is the frequency-to-time mapping; for an input signal with negligible chirp relative to the added dispersion of the optical system, we have the mapping of the envelope angular frequency $\Omega_s = \tau_g^{-1}(t)$ to a corresponding time $t$, given dispersive elements with a total group delay of $\tau_g(\Omega_s)$.

There are primarily two effects of dispersion: a power scaling by a factor of:

$$\alpha(t) = \frac{\tau_g'(\tau_g^{-1}(t))}{\left(\tau_g\left(\frac{\Delta B_{in}}{2}\right) - \tau_g\left(-\frac{\Delta B_{in}}{2}\right)\right)/\Delta B_{in}} \quad (2)$$

and a bandwidth reduction by a factor of:

$$\mu(t) = \frac{B_{PD}}{B_\omega(\Omega_s)} = \frac{4\Delta T_{in}\Delta B_{in}/\left(\tau_g\left(\frac{\Delta B_{in}}{2}\right) - \tau_g\left(-\frac{\Delta B_{in}}{2}\right)\right)}{\sqrt{\frac{1}{4\pi}|\tau_g'(\tau_g^{-1}(t))|}} \quad (3)$$

where we have assumed $\Delta B_{in}$ is the input optical bandwidth, $B_\omega(\omega)$ is the instantaneous frequency of the original signal prior to stretching at time $t$, $\Delta T_{in}$ is the input pulse duration, and $B_{PD}$ is the matched photodetector bandwidth. Tuning the dispersion profile thus enables us to modify the signal SNR. Note that for normalization of both scaling factors, we used an equivalent linear stretch with the same output duration as the warped stretch (i.e. $\beta_2 L = \left(\tau_g\left(\frac{\Delta B_{in}}{2}\right) - \tau_g\left(-\frac{\Delta B_{in}}{2}\right)\right)/\Delta B_{in}$), as illustrated in Figure 1; other normalization methods are also possible.

To capture the time dependent effects of the tailored dispersion, we define the instantaneous SNR of the output signal as: $SNR_i(t) = P_{out}(t)/\sum_j N_j(t)$, where the instantaneous input signal power $P_{out}(t) = i_{out}^2(t)R$ and the noise powers $\{N_j(t)\}$ are understood to be time-averaged over a duration equal to the time resolution of photo detection and analog-to-digital conversion.

We thus consider the SNR output of an optical system after warped stretch as:

$$SNR_i(t) = \frac{i_{out}^2(t)}{i_{th}^2(t) + i_{shot}^2(t) + i_{ASE}^2(t) + i_Q^2(t)} \quad (4)$$

where $i_{th} = \sqrt{4k_B T B_{PD}/R} = 4\sqrt{k_B T C}B_{PD}$ is the thermal noise photo-current, $i_{shot}(t) = \sqrt{2q i_{out}(t) B_{PD}}$ is the shot noise photo-current, $i_{ASE}(t) = 2\sqrt{r_d S_{ASE} i_{out}(t) B_{PD}}$ is the dominant term for the ASE photocurrent, and $i_Q(t) = \|i_{out}(t) - (2^{ENOB})^{-1} \cdot \text{round}[2^{ENOB} i_{out}(t)]\|$ is the quantization noise photocurrent.

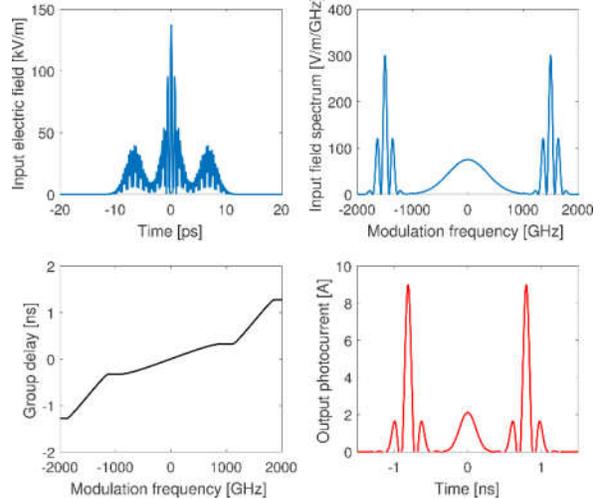

Fig. 3. (a) Electric field of input test signal. (b) Electric field spectrum amplitude of input. (c) Group delay of dispersive element. (d) Amplified output photocurrent in time. Note the occurrence of the warped frequency-to-time mapping between subplots (b) and (d).

The SNR was simulated, taking into thermal noise (T = 300K), shot noise and ASE noise (amplifier gain = 5) of the photocurrent, as well as quantization noise from digitization (ENOB = 8 bits). In the simulation, an optical test signal is passed through a tailored dispersive element, then photodetected and digitized. The input optical power was varied to investigate the effect of the power scaling $\alpha$ on the SNR. For the bandwidth reduction factor $\mu$, both the group delay and the detection bandwidth of the photodetector were simultaneously varied, such that the simulated setup always remains at the optimal detector bandwidth.

## IV. RESULTS AND DISCUSSION

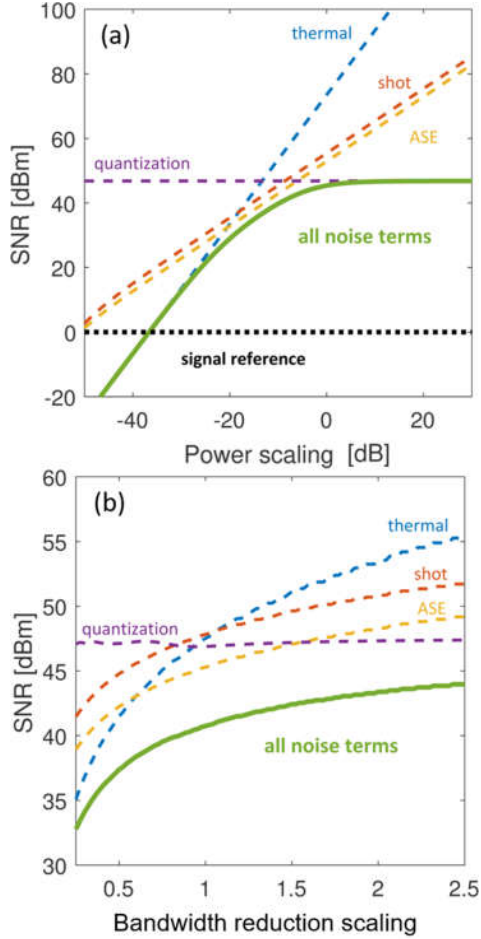

Fig. 3. (a) Power scaling and (b) bandwidth reduction scaling of the instantaneous SNR. The SNR in general has three regimes: the thermal noise-limited regime (SNR scaling by $(\alpha\mu)^2$), the shot and ASE noise-limited regime (SNR scaling by $\alpha\mu$), and the quantization noise-limited regime (SNR constant with input power).

Fig. 2 shows the effect of scaling the input power on the SNR of a warped stretch system. While the instantaneous SNR is limited at high optical powers by the quantization noise, the instantaneous SNR can be improved for weak portions of the signal which are below this threshold. This is achieved by a decrease of dispersion (i.e. increase in $\alpha$) in the corresponding spectral region of the input. Meanwhile, the increase in dispersion (i.e. increase in $\mu$) can improve the SNR by creating oversampled regimes of fast, high power signal portions. Noisy weak signals can thus be SNR-boosted, as long as prior knowledge exists of the input spectrum.

The SNR can be considered in three regimes: the thermal noise-limited regime occurs at low signal powers and high bandwidths (where the SNR scales by $(\alpha\mu)^2$). At moderate stretch factors, we encounter the shot and ASE noise-limited regime (SNR scaling by $\alpha\mu$), and, finally, the quantization noise-limited regime (where SNR is constant with input power).

These results also show the potential for optical encryption by engineering a dispersive element according to the signal's time-averaged spectral characteristics. A broadband signal can be buried under noise floor upon transmission. Upon reception, the inverse dispersion is used to reproduce the optical data.

## V. CONCLUSIONS

Given prior knowledge of a signal's spectral statistics, its SNR can be manipulated (boosted or lowered) in a reversible way. This has applications to weak signal detection and in cybersecurity.

## VI. ACKNOWLEDGEMENTS

This work was supported by the Office of Naval Research under Award Number N00014-14-1-0505 (ONR MURI on Optical Computing).